\documentstyle[preprint,aps,prb]{revtex}

\begin{document}

\draft

\title{Low-temperature kinetics of exciton-exciton annihilation
of weakly localized one-dimensional Frenkel excitons}

\author{I.V.Ryzhov$^1$, G.G. Kozlov$^2$, V.A.Malyshev$^2$, and
J. Knoester$^3$}

\address{$^1$Herzen Pedagogical University, Moika 48, 191186
Saint-Petersburg, Russia}

\address{$^2$National Research Center "Vavilov State Optical Institute",
Birzhevaya Liniya 12, 199034 Saint-Petersburg, Russia}

\address{$^3$Institute for Theoretical Physics and  Material Science
Center, University of Groningen, Nijenborgh 4, 9747 AG Groningen
The Netherlands}

\date{\today}

\maketitle

\begin{abstract}

We present results of numerical simulations of the kinetics of
exciton-exciton annihilation of weakly localized one-dimensional
Frenkel excitons at low temperatures. We find that the kinetics is
represented by two well-distinguished components: a fast
short-time decay and a very slow long-time tail. The former arises
from excitons that initially reside in states belonging to the
same localization segment of the chain, while the slow component
is caused by excitons created on different localization segments.
We show that the usual bi-molecular theory fails in the
description of the behavior found. We also present a qualitative
analytical explanation of the non-exponential behavior observed in
both the short- and the long-time decay components.  Finally, it
is shown that our theoretical estimate for the annihilation time of
the fast component is in good agreement with data obtained from
transient absorption experiments on J-aggregates of
pseudo-isocyanine.

\end{abstract}


%\narrowtext

\section{Introduction}
\label{Intro}

Exciton-exciton annihilation is an important process that strongly
influences the optical and opto-electronic properties of materials
at high excitation densities. In particular, exciton-exciton
annihilation affects the nonlinear and lasing properties of
organic systems, such as J-aggregates and polymer
films.~\cite{Denton97} Though the importance of this process is
well-recognized, the microscopic understanding of the annihilation
kinetics is still rather poor. This holds especially under the
conditions of strong exciton delocalization and (or) low
temperature, where the usual bi-molecular theory of
exciton-exciton annihilation is expected to break down. The aim of
this paper is to study the annihilation kinetics in weakly
disordered one-dimensional Frenkel exciton systems, where the
exciton coherence size can be considerable (tens of lattice
units). This study is of relevance to the optical properties and
exciton dynamics in J-aggregates and molecular antenna systems.

The standard approach to describe the kinetics of exciton-exciton
annihilation relies on the bi-molecular rate equation, in which it
is assumed that the effective annihilation rate is proportional to
the exciton density. This equation reads~\cite{Stiel88,Sundstrom88,%
vanBurgel95,Sundstrom96,Gadonas97}
\begin{equation}
{\dot n} = -\gamma n - \alpha n^2 \ , \label{BME}
\end{equation}
where $n$ represents the average exciton density, understood here
as the number of excitations per molecule, $\gamma$ is the
single-excitation (radiative and non-radiative) relaxation rate,
and $\alpha$ is the co-called annihilation constant having here
the dimension of 1/time. The effective rate of exciton-exciton
annihilation in the system is then indeed proportional to the
average exciton density $n$ and is given by $\alpha n$. The
solution to Eq.~(\ref{BME}) reads
\begin{equation}
n = {\gamma n_0 \over  \gamma e^{\gamma t} +  \alpha n_0
\left(e^{\gamma t} - 1 \right)} \ , \label{gen_solution}
\end{equation}
where $n_0$ is the initial population of excitations. If the
annihilation dominates the single-exciton relaxation ($\alpha n_0
\gg \gamma$) the excitation population decreases according to a
hyperbolic (non-exponential) law
\begin{equation}
n = {n_0 \over  1 +  \alpha n_0 t} \ . \label{part_solution}
\end{equation}

The typical picture that one commonly has in mind when modeling
the annihilation process as is done in Eq.~(\ref{BME}), is as
follows. First, it is usually understood that, as a result of
strong disorder and (or) high temperature, the Frenkel excitons
represent, in fact, molecular
excitations.~\cite{Stiel88,Sundstrom88,vanBurgel95,Sundstrom96,%
Gadonas97} Next, it is assumed that the excitations (diffusively)
move over the system. If the diffusion rate is large compared to
the rate of nearest-neighbor annihilation, denoted by $w_0$, two
excitations annihilate each other (by fusing into one high-lying
molecular excitation that quickly loses its energy by vibrational
relaxation) when they have reached neighboring molecules. In this
case $\alpha = w_0$. On the other hand, if $w_0$ dominates the
diffusion rate, the annihilation event may occur at a distance
large compared to the the nearest-neighbor separation. In this
case, the annihilation constant is determined by a convolution of
the annihilation rate with the pair correlation function of two
excitations,~\cite{Valkunas95} and may in principle depend on
time.

The bi-molecular rate description has limitations, which become
rather important at low temperatures. Obviously, the bi-molecular
approach does not account for the fact that the excitons in
J-aggregates are generally quite strongly spread, with typical
coherence lengths of several tens of
molecules.~\cite{deBoer90,Fidder90,Fidder91a,Fidder93,Minoshima94,%
Moll95,Kobayashi96} Even at room temperature, this length is of
the order of 10 molecules,~\cite{Bogdanov91,Wang91,Akins97} owing
to the large inter-molecular excitation transfer interaction in
these systems. It is an open and interesting question whether this
finite length may be accounted for by interpreting $\alpha$ in
Eq.~(\ref{BME}) as an effective annihilation constant. In this
paper, we will address this question for low-temperature exciton
systems.

We will study the kinetics of exciton-exciton annihilation of
one-dimensional excitons that are weakly localized by static
disorder. We will use the same framework as was done in
Refs.~\onlinecite{Malyshev99a,Malyshev00a} to calculate
annihilation rates. The new element of the present paper is to use
these rates to follow the kinetics of annihilation. An important
step in describing the annihilation of extended excitons, is to
distinguish between inter- and intra-segment
annihilation.~\cite{Malyshev99a,Malyshev00a} The rationale for
this distinction is as follows. As appears from numerical
simulations of disordered exciton chains, the exciton states
residing close to the bottom of the exciton band (the region that
dominates the optical response) can be classified into groups of a
few (two or three) states. The states within each separate group
are all localized on the same segment of the aggregate, with a
typical size $N^*$ (often referred to as the number of coherently
bound molecules), while the segments corresponding to different
groups do not overlap. In fact, it turns out that the two or three
exciton states within each such group are very similar in
structure and energy to the lowest two or three states that exists
on an ordered chain of length
$N^*$.~\cite{Malyshev91,Malyshev95,Shimizu98,Malyshev99b} In
particular, the lowest state of such a group has a wave function
spread over the segment without nodes and can be interpreted as
the local ground state. The next higher lying state of the group
has a well-defined node and looks like a first local excited
state, etc. The energy difference of the local ground and first
excited states agrees well with that of a perfect chain of length
$N^*$.

Obviously, to describe exciton-exciton annihilation, one should
consider at least the two-exciton states. As is well known,
one-dimensional Frenkel excitons are weakly interacting fermions
(see
Refs.~\onlinecite{Chesnut63,Avetisyan85,Juzeliunas88,Spano91,Spano94}).
Thus, the wave functions of states with two excitons can be
composed of Slater determinants of two one-exciton wave functions.
Under the condition of weak localization, two different types of
two-exciton states then appear: (i) those with two excitons
belonging to the same localization segment, and (ii) those with
the two excitons localized on different segments. This immediately
leads to the distinction of intra-segment and inter-segment
annihilation as fundamentally different annihilation
channel.~\cite{Malyshev99a,Malyshev00a}

This paper is organized as follows. In Sec.~\ref{Model}, we
present our microscopic model of annihilation, express the
annihilation rate in terms of the basic interactions and wave
functions, and make the formal step towards the annihilation
kinetics at low temperatures, where the diffusive motion of
excitons towards each other may be neglected. In Sec.~\ref{Qualit}
we use the distinction between inter- and intra-segment
annihilation to derive qualitatively analytical expressions  for
the low-temperature annihilation kinetics. A more detailed study
is presented in Sec.~\ref{Simul}, where we basically  exactly
solve the kinetics, formally defined in Sec.~\ref{Model}, through
numerical simulations. In Sec.~\ref{Summ} we summarize our
findings and discuss the relevance to experimental low-temperature
annihilation data.

\section{Two-exciton annihilation model}
\label{Model}

\subsection{Motivation}
\label{Motiv}

Under usual experimental conditions,  only a small part of the
localization segments on molecular aggregates are excited. For
example, the authors of Ref.~\onlinecite{Minoshima94,Kobayashi96}
estimated that in their experiments, one hundred molecules per
aggregate were produced at the highest excitation power applied
($0.98GW/cm^2$). As a physical aggregate normally consists of
$\sim 10^4$ molecules,~\cite{Sundstrom88,Sundstrom96} while the
typical localization segment in their particular case counted 20
molecules, these authors concluded that less than one exciton was
created per segment of localization (on average, one exciton per
five segments). A simple consideration based on the Poisson
distribution for the probability of finding an integer number of
excitons per segment, shows that on the physical aggregate about
80 out of 500 segments are expected to be singly excited, while
only 8 are doubly excited. Triply (and more) excited segments are
almost absent. Bearing in mind that excitons, created on the same
segment of an aggregate or on closely spaced separate segments,
will annihilate first, we conclude that a two-exciton model of
annihilation seems to be quite reasonable as a first step.

\subsection{Rate of two-exciton annihilation}
\label{Rate}

As a working model, we adopt a linear chain of ${\cal N}$
three-level molecules as depicted in
Fig.~\ref{fig1}.~\cite{Stiel88,Malyshev99a,Malyshev00a} The two
lower molecular states, denoted "0" and "1", are assumed to form
an exciton band, as a result of a sufficiently strong resonant
dipole-dipole inter-molecular coupling. The corresponding exciton
Hamiltonian, taken in the nearest-neighbor approximation, reads
\begin{equation}
H_{ex} = - U \sum_{n=1}^{{\cal N}-1}(b_{1n}^{+}b_{1,n+1}
+ b_{1,n+1}^{+}b_{1n}) + \sum_{n=1}^{\cal N} E_{1n} b_{1n}^{+}b_{1n}
\ ,
\label{H_ex}
\end{equation}
where $-U < 0$ is the nearest-neighbor hopping integral chosen to
be negative, as is the case for J-aggregates, and $b_{1n}^+
(b_{1n})$ denotes the Pauli creation (annihilation) operator of
the first excited state of  molecule $n$ (the state with all the
molecules in their ground states serves as the vacuum state
$|0\rangle$ and has zero energy). The second term in
Eq.~(\ref{H_ex}) represents the Hamiltonian of non-interacting
molecules, in which $E_{1n} = E_1 + \Delta_n$ is the energy of the
first excited state of molecule $n$ with $E_1$ and $\Delta_n$
being, respectively, the mean value of the energy and a static
random offset. The latter simulates on-site (diagonal) disorder
and results in localization of the excitonic states. Fluctuations
of the nearest-neighbor coupling are neglected. It will be assumed
that $\Delta_n$ is distributed uniformly within the interval
$[-\Delta,\Delta]$, so that the typical magnitude of the disorder
is given  by the standard deviation $\sigma = \Delta/\sqrt 3$. For
$\sigma \ll U$ the exciton eigenfunctions are localized within
rather large segments of the chain: $1 \ll N^* \ll
{\cal N}$.~\cite{Malyshev91,Malyshev95,Shimizu98,Malyshev99b,%
Fidder91b,Bakalis99} Throughout this paper, we will assume that
this condition holds.

The high-lying electronic-vibrational molecular term, depicted as
"2" in Fig.~\ref{fig1},  serves as the intermediate state through
which annihilation occurs.~\cite{Stiel88,Malyshev99a,Malyshev00a}
We consider one of the electronic-vibrational levels to be
resonant with the two-exciton optical states and to undergo an
efficient phonon-assisted relaxation to the ground vibronic state
(see Fig.~\ref{fig1}). The annihilation process itself consists of
transferring the energy of two excitons to the high-lying
molecular term. Assuming this step to occur due to the resonant
dipole-dipole inter-molecular interaction, we may write the
corresponding Hamiltonian as follows
\begin{equation}
H_{a} = {1\over 2} \sum_{m,n=1}^{\cal N} {V\over |n - m|^3}
b_{1n}b_{1m}(b_{2n}^+ + b_{2m}^+) + h.c. \ , \label{Hex-ex}
\end{equation}
where $V$ is the matrix element of the annihilation operator for
nearest neighbors and $b_{2n}^+ (b_{2n})$ denotes the Pauli
creation (annihilation) operator of the high-lying state of
molecule $n$. The operator~(\ref{Hex-ex}) annihilates the two
excitations occupying molecules $m$ and $n$ and excites one of
these molecules in the high-lying state. The implication of $H_a$
for third-order nonlinear optics of J-aggregates has been studied
in Ref.~\onlinecite{Knoester95}.

In accordance with our arguments in Sec.~\ref{Motiv}, we will
assume that not more than two excitons are created by the pump per
linear chain. Moreover, we will assume that $|V|$ is small
compared to the rate, $\Gamma$, of phonon-assisted relaxation in
the high-lying molecular state. We may then use perturbation
theory to calculate the rate of annihilation. Moreover, the back
process (exciton fission) can then be neglected. The resulting
expression for the rate of exciton-exciton annihilation starting
from the two exciton eigenstate $|\mu\nu \rangle$ \ ($1 \le \mu <
\nu \le {\cal N}$) is simply given by the "golden rule"
\begin{equation}
w_{a}^{\mu\nu} = {2\pi \over \hbar} \rho(E_f) \sum_{n=1}^{\cal N}
|\langle 2n|H_{a}|\mu\nu\rangle |^2 \ , \label{Wex-ex}
\end{equation}
where $\rho(E_f)$ is the density of final states (hereafter
replaced  by $1/\Gamma$). Because of the fermionic nature of
one-dimensional Frenkel excitons, one can compose the two-exciton
eigenfunctions as Slater determinants of the one-exciton
eigenfunctions:
\begin{mathletters}
\label{one}
\begin{equation}
|\mu\nu\rangle = \sum_{m=1}^{\cal N} \sum_{n<m}^{\cal N}\psi_{\mu
m;\nu n} |1m,1n\rangle \ , \label{munu}
\end{equation}
\begin{equation}
\psi_{\mu m;\nu n}= \varphi_{\mu m}\varphi_{\nu n}
- \varphi_{\mu n}\varphi_{\nu m} \ ,
\label{psi}
\end{equation}
\end{mathletters}
where $\{\varphi_{\nu n}\}$ are the eigenfunctions of the
one-exciton problem
\begin{equation}
\sum_{m=1}^{\cal N} H_{ex}^{nm}\varphi_{\nu m} = E_\nu
\varphi_{\nu n} \ . \label{EigFunc}
\end{equation}
Here, $H_{ex}^{nm} = \langle 1n| H_{ex} |1m\rangle$ and $E_\nu$ is
the eigenenergy of the one-exciton state $\nu$. Substituting
Eq.~(\ref{psi}) into Eq.~(\ref{Wex-ex}) one
obtains~\cite{Malyshev00a}
\begin{equation}
w_{a}^{\mu\nu} = {2\pi V^2\over\hbar\Gamma} \sum_{m=1}^{\cal N}
\Biggl[{\sum_{n=1}^{\cal N}}\, ^\prime {\psi_{\mu m;\nu n} \over
(m - n)^3} \Biggr]^2 \ , \label{Wex-ex_gen}
\end{equation}
where the prime denotes that $n \ne m$. In particular, for a dimer
$({\cal N} = 2)$ only one two-exciton state exists and its
annihilation rate is given by
\begin{equation}
w_{a}^{12} = w_0 = {4\pi V^2\over\hbar\Gamma} \ . \label{w_0}
\end{equation}

In order to arrive at Eq.~(\ref{Wex-ex_gen}) we used the fact that
Frenkel excitons are non-interacting fermions whenever the
nearest-neighbor approximation is used for the hopping integrals.
They become interacting quasi-particles when including the
coupling to far neighbors. The importance of the latter can be
estimated through the changes which the long-range terms produce
in the density of exciton states.  It is known that in
one-dimensional aggregates, the long-range dipole-dipole
interactions shift the exciton band bottom  by approximately 20\%
compared to the nearest-neighbor
model.~\cite{Malyshev95,Fidder91b} The smallness of this shift
suggests the corrections due to long-range interactions to be of a
perturbative nature. Indeed, for few-particle states, Frenkel
excitons are weakly interacting (well-defined) fermions, despite
the long-range coupling.

Furthermore, as follows from the results of both numerical
simulations~\cite{Fidder91b} and theoretical
estimates~\cite{Malyshev95} of the linear optical properties of
disordered Frenkel chains, the oscillator strengths of the optical
transitions near the lower band edge grow by approximately a
factor of 2.5 due to the long-range dipole-dipole interactions.
This results from a larger extension of the optically active
exciton states in the exact dipole-dipole model compared to the
nearest-neighbor model (at a fixed disorder strength). In
principle, this effect is not of a perturbative nature.  It can,
however, be included into the final formulas for the annihilation
rates (see below) by rescaling the number of coherently bound
molecules $N^*$ . The above arguments justify  the
nearest-neighbor framework as a reasonable approach to describe
Frenkel excitons.

It is worth stressing, though, that the dipole-dipole interaction
in the annihilation  channel ($H_{a}$) can generally not be taken
in the nearest-neighbor approximation, because the annihilation of
two excitons localized on separate localization segments is
determined by the coupling to far neighbors.

\subsection{Low-temperature annihilation kinetics}
\label{Kinetic}

As in the present paper we are mainly interested in the
annihilation kinetics itself, we will neglect any other possible
channel of population relaxation, such as the radiative
transitions from two-exciton to one-exciton states (also acting
towards lowering the exciton population), as well as a possible
multi-phonon relaxation from the high-lying term to the
one-exciton states (acting, on the contrary, towards raising again
the exciton population). To calculate the kinetics of the
exciton-exciton annihilation, we will assume that excitons are
created by a resonant laser pulse, that is short compared to the
inverse of the J-band width. Under these conditions, the initial
populations of the two-exciton state $|\mu\nu\rangle$ is
proportional to the corresponding oscillator strength $F_{\mu\nu}$
given by
\begin{equation}
F_{\mu\nu} = \Big|\langle \mu\nu | D^2 |0\rangle \Big|^2 =
\left(\sum_{m,n=1}^{\cal N} \psi_{\mu m;\nu n}\right)^2 \ ,
\label{F}
\end{equation}
where $D = \sum_{n=1}^N (b_{1n}^{+} + b_{1n})$ is the chain's
dipole operator (the chain length is assumed to be smaller than
the emission wavelength).

After the initial creation process, excitons may in principle move
over the chain,  At low temperature, however, the possibility to
move is very restricted and the optically excited localized
Frenkel excitons are practically immobile.~\cite{Malyshev99a} The
reason is that at low temperature ($T \le $ width of J-band), an
exciton created in one of the local ground states may move to an
other similar state only when the latter has an energy lower than
the former. The typical energy offset between the local ground
states is of the order of the width of their energy distribution
(i.e., the width of the J-band). Therefore, after one jump the
exciton typically resides in the tail of this distribution. The
number of states with still lower energy then drastically reduces,
giving rise to a strong  increase of the mean distance to such
lower energy states. In fact, already after one jump the exciton
has a strongly suppressed chance to jump further, i.e., such a
type of the spatio-energetic diffusion (towards lowering the
energy) is stopped rapidly and does not yield a sufficient
possibility  for two excitons to approach each other and
annihilate. It is worth noting that experiments also indicate the
absence of such a diffusion, which would manifest itself in a red
shift of the exciton emission spectrum relative to the absorption
spectrum. The experimental data show that such a Stokes shift is
either absent or has a small
magnitude.~\cite{deBoer90,Fidder90,Moll95} A similar situation
occurs in glasses doped with rare-earth ions.~\cite{Basiev87}

Following the above arguments, we will assume that two excitons
annihilate from the positions where they have been created. Then,
the time dependence of the population of the excited two-exciton
states is given by
\begin{equation}
P_a(t) = 2\left\langle \sum_{\mu\nu} f_{\mu\nu} \exp(-
w_a^{\mu\nu} t) \right\rangle \ , \label{P_a}
\end{equation}
where $f_{\mu\nu}=F_{\mu\nu}/\sum_{\mu\nu} F_{\mu\nu}$ and angular
brackets denote an average over the disorder realizations. Note
that we have normalized the population such that it equals 2 at $t
= 0$. The formula~(\ref{P_a}) will be the basis of our further
analysis of the low-temperature annihilation kinetics.

\section{Qualitative picture}
\label{Qualit}

Before carrying out numerical simulations, we first provide a
qualitative analysis of Eq.~(\ref{P_a}). Following the arguments
concerning the nature of the low-energy weakly localized
one-dimensional states, we separate the summation in
Eq.~(\ref{P_a}) into two parts:  $P_a(t) = P_a^{intra}(t) +
P_a^{inter}(t)$. The first part,  $P_a^{intra}(t)$, includes all
those terms $\{\mu\nu\}$ where the one-exciton states $\mu$ and
$\nu$ are localized on the same chain segment (doubly excited
segments). The second part,  $P_a^{inter}(t)$, contains those
terms where  $\mu$ and $\nu$ reside on different segments. The
fact that this distinction can only be made for low-energy states
is no restriction, as anyhow these states are the ones that
dominate the ground state to one-exciton and the
one-to-two-exciton absorption spectrum. Using the picture of
exciton states on a chain of effective length $N^*$, one arrives
at the intra-segment annihilation
rate:~\cite{Malyshev99a,Malyshev00a}
\begin{equation}
w_a^{intra} = {5\pi^6\over 18(N^* + 1)^3}\, w_0 \ ,
\label{w_intra}
\end{equation}
where $w_0$ is given by Eq.~(\ref{w_0}) and the factor $5\pi^6/18
\approx 270$.

The second term, $P_a^{inter}(t)$, governs the annihilation of two
excitons created on different localization segments and is
characterized by the rate~\cite{Malyshev00a}
\begin{equation}
w_a^{inter} = {N^* + 1 \over R^6}\, w_0 \ , \label{w_inter}
\end{equation}
where $R$ is the distance between the two excited segments. Note
that the rate of the inter-segment annihilation scales linearly
with $N^* + 1$. As the wave functions of both segments enter the
expression for $w_a^{inter}$, one might intuitively expect a
quadratic dependence on $N^* + 1$. However, only one of the two
excited segments, namely the one that passes to the ground state
$|0\rangle$ in the annihilation process, coherently contributes to
$w_a^{inter}$ giving the factor $N^* + 1$  (so-called superradiant
transition). The transition within the other excited segment
occurs to the high-lying molecular state of each molecule. It is
important to note that the latter events are summed incoherently,
as is evident from Eq.~(\ref{Wex-ex_gen}), thus preventing the
appearance of an extra power of $N^* + 1$.

Keeping in mind that $R \ge N^*$ as well as that the numerical
factor in Eq.~(\ref{w_intra}) is fairly large, one immediately
deduces from Eqs.~(\ref{w_intra}) and~(\ref{w_inter}) that
$w_{a}^{intra} \gg w_{a}^{inter}$ provided that $N^*$ is of the
order of or larger than several units, which is the condition we
will focus on in the simulations. From this, one expects that the
kinetics of the exciton-exciton annihilation will consist of two
distinct parts: a very fast short-time decay, described by
$P_a^{intra}(t)$, and a very slow long-time tail for which
$P_a^{inter}(t)$ is responsible. In order to estimate the relative
weight of these two components, we will take into account the fact
that the oscillator strengths of double excitation of a particular
localization segment and excitation of two different segments are
of the same order. Then, statistical arguments based on the
Poisson distribution seem to be sufficient for making estimates.
If the initial density of excitations equals $n_0$ (in our case,
$n_0 = 2/{\cal N}$), the probability of finding a typical
localization segment (of length $N^*$) to be $k$-fold excited
reads
\begin{equation}
p(k) = {(n_0N^*)^k \over k!} e^{-n_0N^*} \ . \label{Poisson}
\end{equation}
Therefore, the probabilities of double excitation of a typical
localization segment and of excitation of two different segments
are given by\ \ $0.5 \big(n_0N^*\big)^2 e^{-n_0N^*}$  and\ \ $2
n_0N^* e^{-n_0N^*}$, respectively. Thus, the relative contribution
of the shorter component in $P_a(t)$ is of the order of\ \ $0.25
n_0N^*$, which drops upon increasing the disorder strength
$\sigma$ (or upon a corresponding decrease of $N^*$) and grows
with increasing the density  of excitations $n_0$.

Now, we turn to a discussion of the character of the annihilation
kinetics. We recall that in the bi-molecular model, the decay is
non-exponential, as a result of the nonlinearity of the driving
equation~(\ref{BME}). We will argue that in our case, the kinetics
is also non-exponential, which, however, does not result from a
nonlinearity. Let us consider first the intra-segment channel of
annihilation. Here, according to Eq.~(\ref{P_a}), the
nonexponentiality is expected even in the absence of disorder
(i.e., for a regular chain), because the annihilation rate
$w_a^{\mu\nu}$ obviously depends on which exciton states are
involved in the annihilation process (see the next Section). In
the presence of disorder, there is an additional source for the
non-exponential behavior of annihilation through the intra-segment
channel. It originates from the fluctuations in the sizes of the
localization segments, $N$. Equation~(\ref{w_intra}) gives  the
typical magnitude of the intra-segment annihilation rate. In
reality, $N^*$ in Eq.~(\ref{w_intra}) should be replaced by a
fluctuating value $N$. Consequently, the sum over the overlapping
states in Eq.~(\ref{P_a}) can be approximately substituted by an
average over a distribution of $N$, $G(N)$. One then obtains
\begin{equation}
P_a^{intra}(t) \approx 2 \int dN G(N) \exp\left[- w_a^{intra}(N)t
\right]\ . \label{Pa_intra}
\end{equation}
As follows from numerical
simulations,\cite{Malyshev95,Shimizu98,Fidder91b,Malyshev00b} the
standard deviation of N, $[\int dN G(N) (N - {\bar N})^2]^{1/2}$,
is of the order of the mean, ${\bar N} = \int dN G(N) N$, i.e.,
the distribution $G(N)$ is rather broad. Due to this fact, the
resulting nonexponentiality is expected to be considerable. The
numerical simulations presented in Sec.~\ref{Simul} confirm this
picture.

The origin of the non-exponential behavior of the inter-segment
annihilation is two-fold. First, the corresponding annihilation
rate, as in the previous case, depends on the size of the
localization segment $N$ (see Eq.~(\ref{w_inter}) ). Thus,
fluctuations of the latter will affect the annihilation kinetics
even for a fixed distance between excited segments. However, the
character of the inter-segment annihilation is determined mostly
by the strong dependence  of the annihilation rate (\ref{w_inter})
on the distance between two excited segments, $R$. The
annihilation kinetics caused by the fluctuations of $R$ is simply
given by an average of the pair kinetics $2\exp
[-w_a^{inter}(R)t]$ over all realizations of R. In order to obtain
an analytical estimate, let us assume that the density of excited
segments is low (as is in our case), so that $R$ can be treated as
a continuous stochastic variable. Then the annihilation kinetics
is given by an integral similar to that in Eq.~(\ref{Pa_intra})
with  $G(N)$ replaced by a suitable $R$-distribution function,
$G(R)$. We will adopt a uniform distribution for $R$, $G(R) =
N^*/{\cal N}$, assuming that the probability of finding a segment
to be excited is equal to the inverse of the number of segments in
the chain, ${\cal N}/N^*$. In evaluating the integral, we will
extend the integration over the entire positive axes, neglecting
thus the minimal distance between two adjacent segments as well as
the finiteness of the chain. Both approximations are justified at
a low density of excitations. One thus arrives at
\begin{equation}
P_a^{intra}(t) \approx 2 \left[ 1 - {N^* \over {\cal N}}
\int_0^{\infty} dR \left(1 - e^{- w_a^{inter}(R)t} \right) \right]
\approx 2 \left[ 1 -  \Gamma\left({5 \over 6}\right) {N^* \over
{\cal N}} (N + 1)^{1/6}(w_0t)^{1/6} \right]\ , \label{Pa_inter}
\end{equation}
where $\Gamma(x)$ is the Gamma-function. Equation~(\ref{Pa_inter})
is correct provided that the second term on the right-hand site is
less than unity. This holds in a very large time interval,
determined by the inequality \ $\Gamma\left({5/6}\right) N^* (N +
1)^{1/6} (w_0t)^{1/6} < {\cal N}$. It follows from the stretched
exponential behavior of Eq.~(\ref{Pa_inter}) that further
averaging of $P_a^{intra}(t)$ over the $N$-distribution will not
change the character of the kinetics and results, in fact, in
replacing $N$ by $N^*$.

To conclude this section, we note that the ("artificial") quantity
$P_a(t)$ can be simply rescaled to the measurable magnitude - the
density of excitons $n(t) = P_a(t)/{\cal N}$. Introducing the
initial exciton density $n_0 = 2/{\cal N}$, we obtain
\begin{equation}
n^{intra}(t) \approx n_0 \left[ 1 - {1\over 2} \Gamma \left({5
\over 6}\right) n_0 N^* (N^* + 1)^{1/6}(w_0t)^{1/6} \right]\ ,
\label{n_inter}
\end{equation}

\section{Numerical simulations and discussion}
\label{Simul}

To study the annihilation kinetics in more detail, we have carried
out numerical simulations for a chain of length ${\cal N} = 200$.
For such a length, the mean initial density of excitations is $n_0
= 0.01$.  We calculated the one-exciton eigenfunctions
$\varphi_{\nu n}$ by diagonalizing numerically the Frenkel
Hamiltonian~(\ref{H_ex}) for a particular realization of disorder,
and then composed two-exciton eigenfunctions according to
Eq.~(\ref{one}). Using further Eq.~(\ref{Wex-ex_gen}) and
Eq.~(\ref{F}), we computed the rate of annihilation,
$w_a^{\mu\nu}$, and the oscillator strength, $F_{\mu\nu}$, for any
two-exciton state $|\mu\nu \rangle$. Then Eq.~(\ref{P_a}) was used
to evaluate the annihilation kinetics. The resulting kinetics was
obtained by averaging over 20 realizations of disorder. An
increase of this number did not lead to considerable changes in
the calculated curves. As a time unit, we used $w_0^{-1}$. The
results of the simulations for different values of $\Delta/U$ are
depicted in Figs.~\ref{fig2}-\ref{fig5} by thick solid lines.

Figure~\ref{fig2} represents the annihilation kinetics for a
perfect chain ($\Delta = 0$), as well as a least-square fit by
means of an exponential $2\exp(-w_0t/a)$ (thin solid line),
achieved at $a = 1.63\cdot 10^4$. The fit clearly  demonstrates
that the calculated curve can not be matched by a single
exponential. This unambiguously means that not only the states
with $\mu = 1$ and $\nu = 2$, having the largest oscillator
strengths, contribute to the sum in Eq.~(\ref{P_a}), but the other
states contribute a comparable amount. The time scale of the
kinetics depicted in Fig.~\ref{fig2} qualitatively corresponds to
that calculated by using Eq.~(\ref{w_intra}) with $N^*$ replaced
by ${\cal N}$: at ${\cal N} = 200$, one obtains  $w_a^{intra} \sim
3\times 10^{-5}w_0$.

We also tried to fit the numerical curve in Fig.~\ref{fig2} by
means of the bi-molecular equation~(\ref{part_solution}) taken in
the form $2/(1 + bw_0t)$. The best fit was achieved at $b =
1.07\cdot 10^{-4}$ and is plotted in Fig.~\ref{fig2} by the dashed
line. At first glance, it seems that the latter almost matches the
numerical data except, maybe, at the initial stage. However, the
fitting constant $b$, carrying, in fact, the meaning of the
density of excitations (see the discussion presented in the
Introduction), underestimates  the real value $n_0 = 0.01$ by two
orders of magnitude.

Figures~\ref{fig3} and~\ref{fig4} show the numerical results
obtained for disordered chains with  different degrees of
disorder, $\Delta/U$. In Fig.~\ref{fig3}, we plotted the
annihilation decay curves in a wide time interval, while
Fig.~\ref{fig4} presents the initial stages of the annihilation
process. From the numerical results presented in Figs.~\ref{fig3}
and~\ref{fig4} several conclusions can be deduced. First of all,
it is clearly seen that the entire kinetics indeed consists of two
well-distinguished components: a fast short-time decay, becoming
faster as the disorder is increased, and a very slow long-time
tail. The weight of the faster component is smaller than that for
the slower one and drops upon increasing the disorder strength
$\Delta/U$.

It is reasonable to relate these two components to the intra- and
inter-segment channels of exciton-exciton annihilation,
respectively, in accordance with the qualitative picture discussed
in the previous Section. Indeed, let us use for the typical size
$N^*$ of a localization segment the well-known
estimate~\cite{Malyshev91,Malyshev95,Fidder91b,Bakalis99}
\begin{equation}
N^* + 1 = \left ( 3\pi^2 {U\over\sigma} \right )^{2/3} \ .
\label{N*}
\end{equation}
Recall that in our case $\sigma = \Delta/\sqrt 3$. Substituting
Eq.~(\ref{N*}) into (\ref{w_intra}), one gets
\begin{equation}
w_{a}^{intra} = {5\pi^2 \over\ 506} \left( {\Delta\over
U}\right)^2 w_0 \approx 0.1 \left( {\Delta\over U}\right)^2 w_0 \
, \label{1w_intra}
\end{equation}
Equation~(\ref{1w_intra}) gives us the disorder scaling of the
intra-segment annihilation rate. Accordingly, we arrive at the
following estimates: $w_{a}^{intra} \sim 4\times 10^{-3}w_0,
2\times 10^{-2}w_0$ and $6\times 10^{-2}w_0$ respectively for
$\Delta/U = 0.2, 0.4$ and $0.8$. Indeed, these numbers
qualitatively match the time scales of the fast components (see
Fig.~\ref{fig4}).

We also plotted in Figs.~\ref{fig3} and~\ref{fig4} the least
square fits of the numerical data by means of the function $2 -
c(w_0t)^{1/6}$ (thin solid line). One observes that at higher
degree of disorder $(\Delta/U = 0.8)$, when the weight of the
faster component is smallest, the fitting function fairly well
follows the numerical curve over almost the entire time interval
of decay. The value of the fitting constant $c = 0.14 - 0.15$ is
of the same order as the one deduced from the theory,
Eq.~(\ref{Pa_inter}), according to which it must be
$2\Gamma(5/6)(N^*)^{7/6}/{\cal N} = 2\Gamma(5/6) (3{\sqrt 3}\pi^2
U/\Delta)^{7/9}/{\cal N} \approx 0.24$ (The discrepancy probably
stems from neglecting $N^*$ as a minimal separation in
Eq.~(\ref{Pa_inter}).) This unambiguously means that the
inter-segment channel dominates the long-time part of the
annihilation kinetics. It should be especially stressed that  the
bi-molecular fits, shown in Figs.~\ref{fig3} and~\ref{fig4} by the
dashed lines, fail absolutely in the description of the numerical
data.

In order to show the character of the decay (exponential or
non-exponential) in the case of disordered chains, we depicted in
Fig.~\ref{fig5}  the log-plot of the calculated annihilation
kinetics for $\Delta/U = 0.2$. As can be seen, neither of the two
components shows an exponential behavior.

\section{Summary and concluding remarks}
\label{Summ}

In this paper we have studied the low-temperature kinetics of
exciton-exciton annihilation of weakly localized one-dimensional
Frenkel excitons using a two-exciton static model (immobile
quasi-particles) with diagonal disorder. Our analysis leads to
three main conclusions:

i -   The entire kinetics consists of two well-distinguished
      components: a very slow long-time decay and a much faster
      short-time drop. The latter component becomes faster with
      higher degree of disorder.
      The weight of the faster component is much smaller than
      that of the slower one, and decreases with increasing
      disorder strength.

ii -  Neither of these two components shows an exponential
      behavior.

iii - The usual bi-molecular theory fails in the description of
      the behavior found.

\noindent These findings are well-understood from the existence of
two competing options for two excitons to annihilate. The slower
component is driven by the annihilation of exciton states
localized on different segments of the chain, while the faster one
originates from the annihilation of doubly excited segments.
Fluctuations of distances between two excitons and sizes of the
localization segments explain the non-exponential nature of the
slower and faster components, respectively.

It is worthwhile to estimate the typical rates of both
annihilation channels for existing J-aggregates. In order to do
this, we need information concerning the parameters $U$, $V$, and
$\Gamma$. For J-aggregates $U \sim 1000$ cm$^{-1}$ is quite
typical.~\cite{deBoer90,Fidder91a,Moll95,Spano94} Less information
consists concerning the annihilation interaction $V$, but as we
assumed it to be of dipolar origin it seems not unreasonable to
take a value similar to $U$. This is, in fact, supported by
semi-empirical calculations of higher molecular singlet states of
pseudo-isocyanine (PIC) molecules.~\cite{Scherer95} These
calculations indicate a molecular $S_1 \rightarrow S_2$ transition
that is similar in energy and oscillator strength as the $S_0
\rightarrow S_1$ transition responsible for PIC's well-known
J-band. We thus take $V \sim 1000$ cm$^{-1}$. Finally, we will
take $\Gamma \sim 3000$ cm$^{-1}$, corresponding to a vibrational
relaxation time in the $S_2$ state of about 10 fs. Using these
numbers, we arrive at $w_a^{intra} \sim 3\times 10^{16}{N^*}^{-3}$
s$^{-1}$. The corresponding estimate for the inter-segment
annihilation rate taken for adjacent segments ($R = N^*$) reads
$w_a^{inter} \sim 10^{14}{N^*}^{-5}$ s$^{-1}$.

At low temperatures, the quantity $N^*$ is found to be of the
order of several
tens~\cite{deBoer90,Fidder91a,Fidder93,Minoshima94,Moll95}.
Letting $N^* = 20$, as was reported in
Refs.~\onlinecite{Minoshima94} and~\onlinecite{Kobayashi96}, we
arrive at $w_a^{intra} \sim 4\times 10^{12}$ s$^{-1}$ and
$w_a^{inter} \sim 3\times 10^7$ s$^{-1}$. Note that the magnitude
of $w_a^{inter}$ appears to be even smaller than the spontaneous
emission rate of a single molecule, which typically is of a few
times $10^8$ s$^{-1}$. Certainly $w_a^{inter}$ is much smaller
than the spontaneous emission rate for an exciton state. It is to
be noted furthermore that raising $N^*$ by a factor of 2 will
reduce $w_a^{inter}$ by almost two orders of magnitude. From the
above, an important conclusion can be deduced: the inter-segment
channel of exciton-exciton annihilation is in fact ineffective at
low temperatures, because the radiative relaxation is much faster.
On the contrary, the intra-segment annihilation rate is fairly
high and should be viewed as the unique way for two weakly
localized excitons to annihilate at low temperature. However,
since this process occurs only for doubly excited localization
segments, it will affect the entire exciton population only if the
number of doubly excited segments is high, i.e., at sufficiently
high laser intensities.

We note that the separation into inter-segment and intra-segment
annihilation channels was in fact concluded from transient
absorption experiments on PIC J-aggregates at
20\,~K.~\cite{Minoshima94,Kobayashi96} The part of the observed
annihilation kinetics with a decay time of 200 fs may indeed be
related to intra-segment annihilation, as is clear from our
estimate for $w_a^{intra}$. Our estimate for $w_a^{inter}$ shows,
however, that it is unlikely that the second component of the
kinetics reported in Refs.~\onlinecite{Minoshima94}
and~\onlinecite{Kobayashi96}, with a decay time of 1.5 ps, may
indeed be ascribed to inter-segment annihilation (see also the
discussion in Refs.~\onlinecite{Malyshev99a}
and~\onlinecite{Malyshev00a}).

Our findings concerning the ineffectiveness of inter-segment
annihilation provide us with a way to control the exciton-exciton
annihilation at low temperature. Indeed, recall that the local
ground and first excited states belonging to the same localization
segment are separated by the energy offset $E_2^* - E_1^* =
3\pi^2U/(N^* + 1)^2$, which is of the order of the J-band
width.~\cite{Malyshev91} For typical J-aggregates, the exciton
radiative rate $\gamma$, representing the unique relaxation
constant at low temperatures, is much smaller than this energy
mismatch. Hence one may get a large number of localization
segments to be singly excited by applying a field with Rabi
frequency smaller than $E_2^* - E_1^*$, but larger than $\gamma$.
At the same time, none of the localization segments will be doubly
excited. Therefore, under such conditions, a fairly large exciton
population may be created in J-aggregates, without being affected
by exciton-exciton annihilation.

\acknowledgments

This work has been partially supported by  INTAS (project No.
97-10434). G.\ G.\ K.\ and V.\ A.\ M.\ also acknowledge  support
from  the German Federal Ministry of Education, Science, Research,
and Technology within the TRANSFORM - programme (project No. 01 BP
820/7). V.\ A.\ M.\ acknowledges the University of Juv\"askyl\"a,
where this work was started, for hospitality.

\begin{figure}
\caption{Schematic representation of all interactions contributing
to the exciton-exciton annihilation process. The interaction
$H_{ex}$, indicated by the dashed lines, forms excitonic states in
the subspace of the molecular states "0" and "1". Excitons
annihilate through a high-lying electronic-vibrational molecular
term "2" ($\omega_{10} \approx \omega_{21}$). The first step of
the annihilation process results from the inter-molecular
interaction $H_a$, which induces  simultaneous transitions of the
molecule $m$ to the ground state and molecule $n$ to the
high-lying term.  The second step results from fast vibrational
relaxation within high-lying electronic-vibrational sub-levels
towards the ground vibrational state characterized by a rate
$\Gamma\gg H_a$.} \label{fig1}
\end{figure}

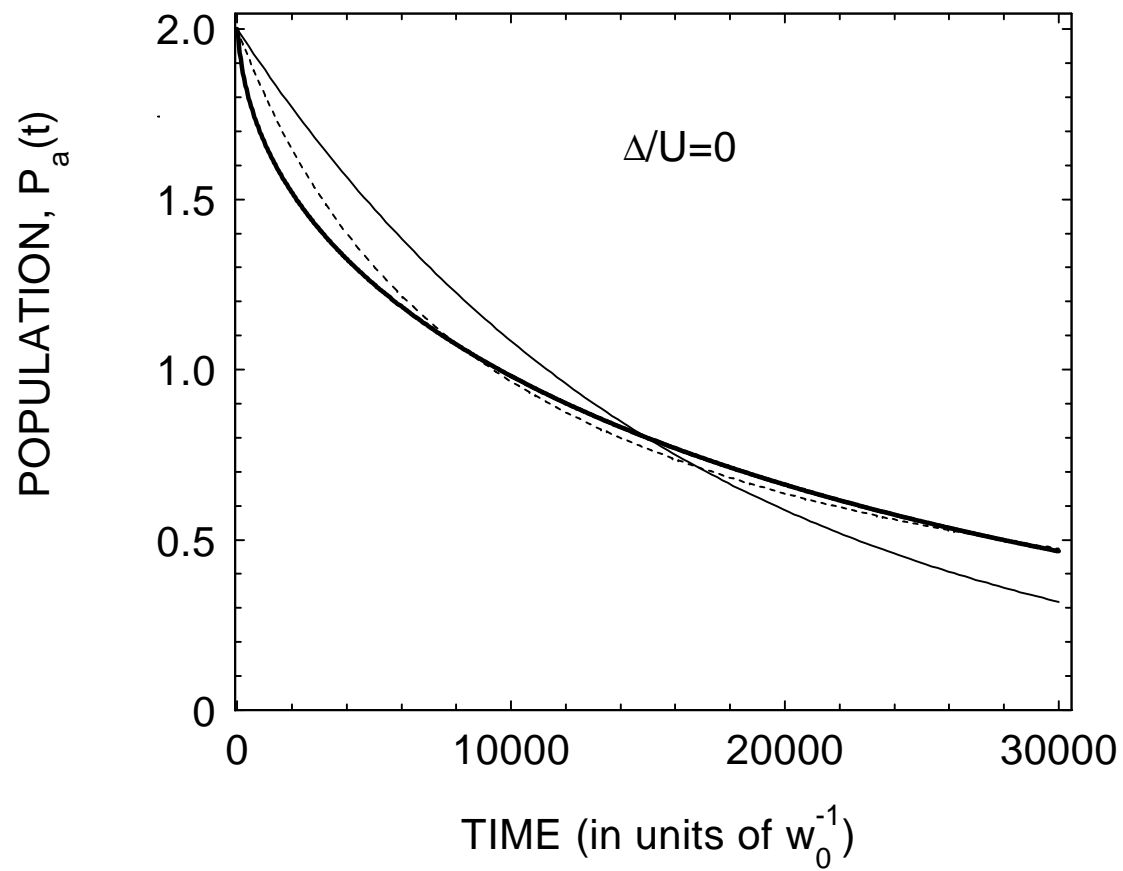
\begin{figure}
\caption{Plot of the exciton-exciton annihilation kinetics
obtained from numerical simulations  for a regular linear chain of
200 sites (thick solid line). The least-square fits by means of
the exponential $2\exp(-w_0t/a)$  with $a = 1.63\cdot 10^4$  as
well as by the bi-molecular model~(\protect\ref{part_solution}),
taken in the form $2/(1 + bw_0t)$ with $b = 1.07\cdot 10^{-4}$ are
presented by the thin solid  and dashed lines, respectively. The
time unit is chosen to be $w_0^{-1}$ (see
Eq.(\protect\ref{w_0})).} \label{fig2}
\end{figure}

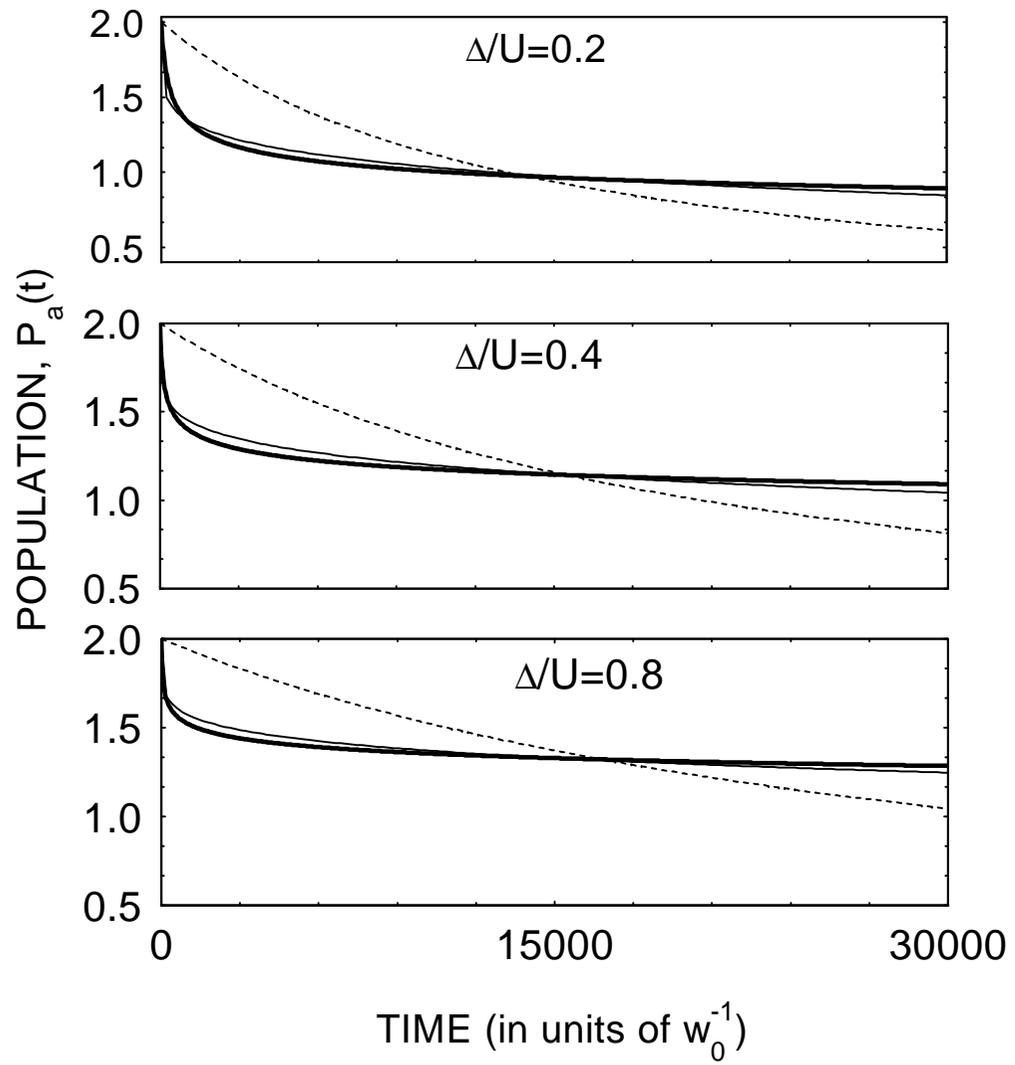
\begin{figure}
\caption{Plots of the exciton-exciton annihilation kinetics
obtained from numerical simulations (thick solid lines) for a
linear chain of 200 sites at different values of the degree of
disorder $\Delta/U$. Thin solid lines give the least-square fits
by means of the function $2 - ct^{1/6}$ at  $c = 0.21\ (\Delta/U =
0.2)$,  $c = 0.17\ (\Delta/U = 0.4)$ and $c = 0.14\ (\Delta/U =
0.8)$. Dashed lines give the best fit using  the bi-molecular
model~(\protect\ref{part_solution}). The time unit is chosen to be
$w_0^{-1}$.} \label{fig3}
\end{figure}

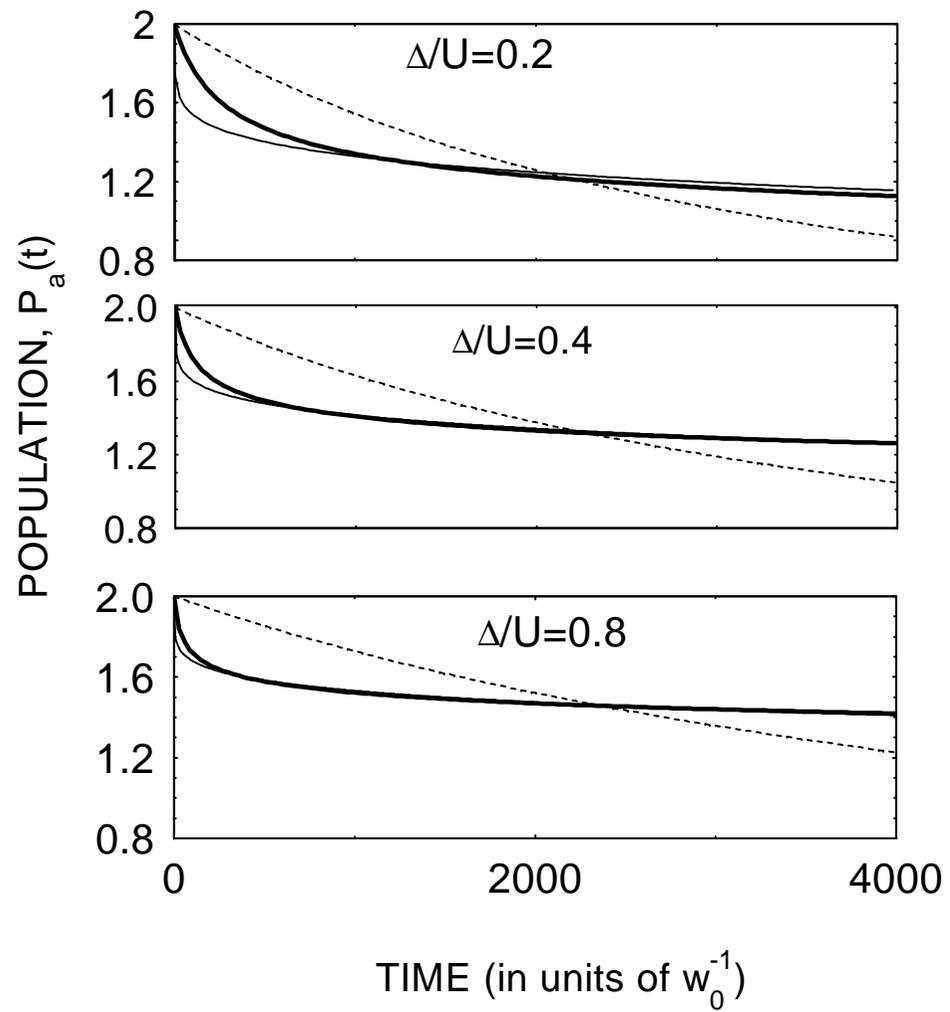
\begin{figure}
\caption{As Fig.~\protect\ref{fig3} but now focused on the initial
stage of the annihilation process.  For this time interval, the
best fits by means of the function $2 - ct^{1/6}$ were achieved at
$c = 0.21\ (\Delta/U = 0.2)$,  $c = 0.18\ (\Delta/U = 0.4)$ and $c
= 0.15\ (\Delta/U = 0.8)$. } \label{fig4}
\end{figure}

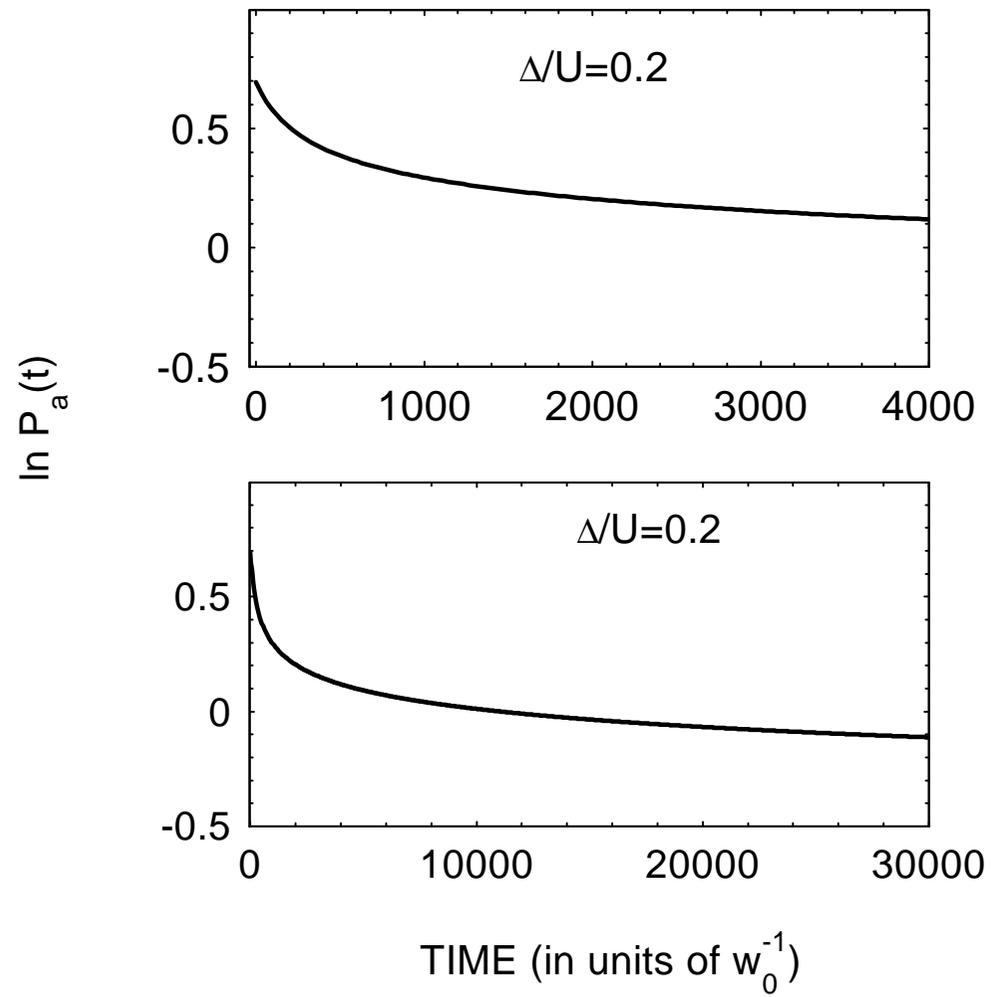
\begin{figure}
\caption{Log-plot of the exciton-exciton annihilation kinetics for
a linear chain of 200 sites at $\Delta/U = 0.2$, demonstrating the
non-exponential character of the decay. The time unit is chosen to
be $w_0^{-1}$.} \label{fig5}
\end{figure}

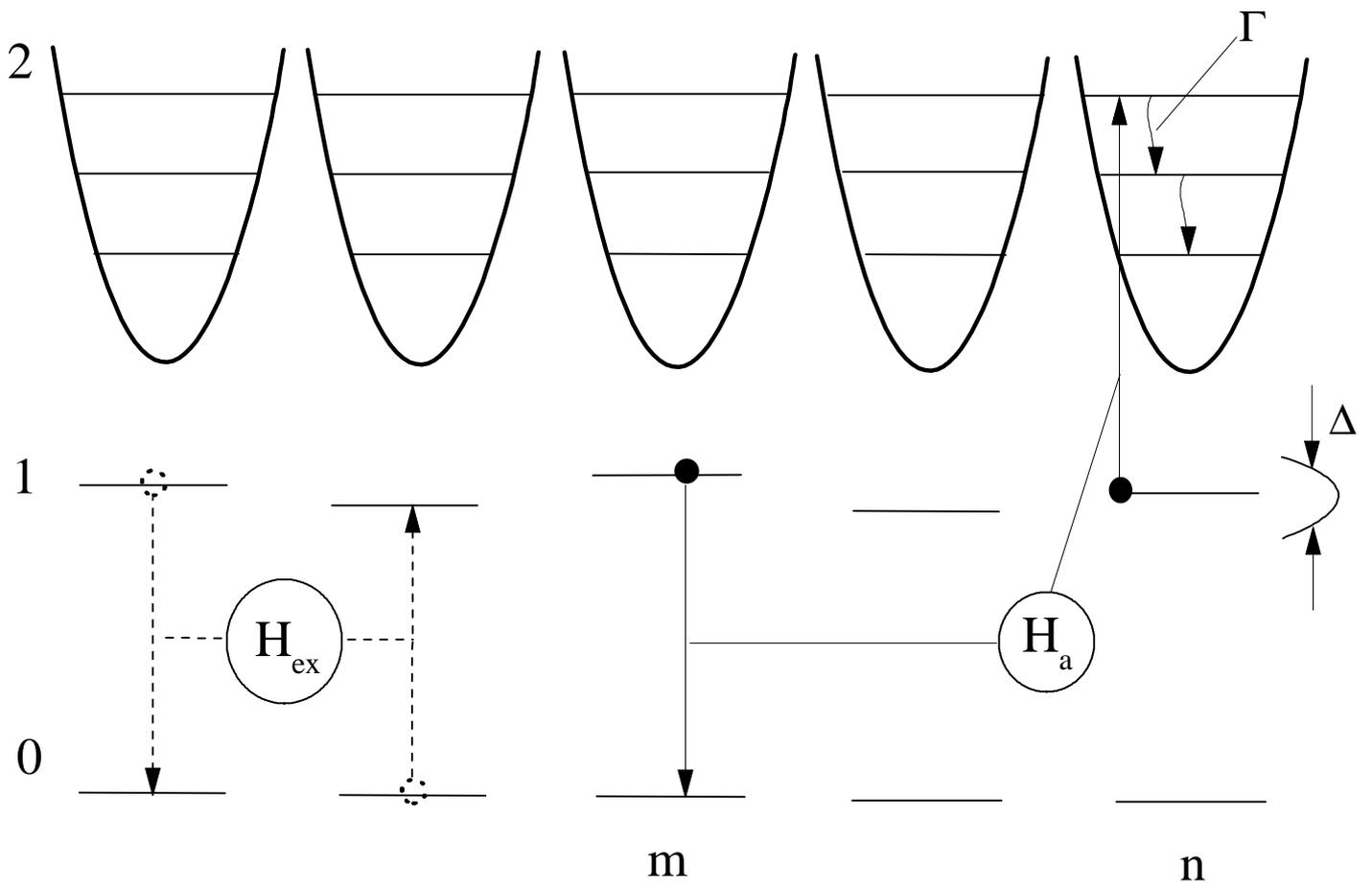
\end{document}